\input harvmac

\def\frac#1#2{{#1\over #2}}

\def\IZ{\relax\ifmmode\mathchoice
{\hbox{\cmss Z\kern-.4em Z}}{\hbox{\cmss Z\kern-.4em Z}}
{\lower.9pt\hbox{\cmsss Z\kern-.4em Z}}
{\lower1.2pt\hbox{\cmsss Z\kern-.4em Z}}\else{\cmss Z\kern-.4em
Z}\fi}
\def\IC{{\relax\,\hbox{$\inbar\kern-.3em{\rm C}$}}}

\font\cmss=cmss10 \font\cmsss=cmss10 at 7pt
\def\IR{\relax{\rm I\kern-.18em R}}

\def\IZ{\relax\ifmmode\mathchoice
{\hbox{\cmss Z\kern-.4em Z}}{\hbox{\cmss Z\kern-.4em Z}}
{\lower.9pt\hbox{\cmsss Z\kern-.4em Z}}
{\lower1.2pt\hbox{\cmsss Z\kern-.4em Z}}\else{\cmss Z\kern-.4em
Z}\fi}
\def\IC{{\relax\,\hbox{$\inbar\kern-.3em{\rm C}$}}}

\def\frac#1#2{{#1\over #2}}

\Title{\vbox{\baselineskip11pt
\hbox{hep-th/0406058}
 \hbox{HUTP-04/A026}
  }}
  {\vbox{ \centerline{Two Dimensional Yang-Mills,}
\vskip .8cm
\centerline{Black Holes}
\vskip .8cm
\centerline{and Topological Strings} \vskip 4pt
}}
\centerline{
Cumrun Vafa}
\medskip
\vskip 8pt

\centerline{\it 
Jefferson Physical Laboratory, Harvard University,}
\centerline{\it Cambridge, MA 02138, USA}
\medskip
\noindent

We show that topological strings on a class of non-compact
Calabi-Yau threefolds is equivalent to two dimensional bosonic $U(N)$ Yang-Mills
on a torus. 
We explain this correspondence using the recent results
on the  equivalence of the partition function of topological strings
and that of four dimensional BPS black holes, which in turn is holographically dual to the field theory on 
a brane.  The partition function of the field theory on the brane
reduces, for the ground state sector, to that of $2d$ Yang-Mills theory.  We conjecture that
the large $N$ chiral factorization of the 2d $U(N)$ Yang-Mills
partition function reflects the existence of two
boundaries of the classical $AdS_2$ geometry, with one chiral
sector associated to each boundary;  moreover the
lack of factorization at finite $N$
 is related to the transformation
of the vacuum state of black hole from a pure state at all orders in $1/N$ to a state which appears mixed at finite 
$N$ (due to $O(e^{-N})$ effects).

\smallskip

\Date{June, 2004}


\newsec{Introduction}

In a recent paper  \ref\osv{H.~Ooguri, A.~Strominger and C.~Vafa,
``Black hole attractors and the topological string,''
arXiv:hep-th/0405146.}, following
earlier work
\ref\LopesCardosoWT{
G.~Lopes Cardoso, B.~de Wit and T.~Mohaupt,
``Corrections to macroscopic supersymmetric black-hole entropy,''
Phys.\ Lett.\ B {\bf 451}, 309 (1999)
[arXiv:hep-th/9812082].
}, a relation was proposed between the partition function
of 4d BPS black holes
and the topological string partition function:
$$Z_{BH}=|Z_{top}|^2.$$
  On the other
hand the black hole partition function is given by the partition function
of a quantum field theory living on the brane which produces the black
hole:
$$Z_{BH}=Z_{brane}$$
which in turn leads to
\eqn\funs{Z_{BH}=Z_{brane}=|Z_{top}|^2.}
In some cases, as was noted in \osv , the relevant brane theory
is given by a topologically twisted maximally supersymmetric Yang-Mills
theory. 
The charge of the magnetic D-branes $p^I$ and the chemical potentials
for the electric D-branes $\phi^I$ fix the homogeneous coordinates on the
moduli of the Calabi-Yau by
$$X^I=p^I+i{\phi^I \over \pi}$$
where in the A-model topological string (in a suitable gauge) we have
the topological string coupling constant $g_s$ given by
$$g_s={4\pi i\over X^0}.$$

The aim of this paper is to give a concrete realization of \funs .
In fact we shall see that an anology pointed out in \osv\ can 
be concretely realized in this context:  It was noted in \osv\
that the relation \funs\ is analogous to the relation between
the large $N$ expansion of the partition function of 2d Yang-Mills
theory on a Riemann surface and the existence of an almost factorized structure
into holomorphic and anti-holomorphic maps to the Riemann surface
\lref\GrossTU{
D.~J.~Gross,
``Two-dimensional QCD as a string theory,''
Nucl.\ Phys.\ B {\bf 400}, 161 (1993)
[arXiv:hep-th/9212149].
}
\lref\GrossHU{
D.~J.~Gross and W.~I.~Taylor,
``Two-dimensional QCD is a string theory,''
Nucl.\ Phys.\ B {\bf 400}, 181 (1993)
[arXiv:hep-th/9301068].
}
\lref\CordesFC{
S.~Cordes, G.~W.~Moore and S.~Ramgoolam,
``Lectures on 2-d Yang-Mills theory, equivariant cohomology and topological
field theories,''
Nucl.\ Phys.\ Proc.\ Suppl.\  {\bf 41}, 184 (1995)
[arXiv:hep-th/9411210].
}
\lref\MooreDK{
G.~W.~Moore,
``2-D Yang-Mills theory and topological field theory,''
arXiv:hep-th/9409044.
}
\lref\StromingerYG{
A.~Strominger,
``AdS(2) quantum gravity and string theory,''
JHEP {\bf 9901}, 007 (1999)
[arXiv:hep-th/9809027].
}
\lref\MaldacenaUZ{
J.~M.~Maldacena, J.~Michelson and A.~Strominger,
``Anti-de Sitter fragmentation,''
JHEP {\bf 9902}, 011 (1999)
[arXiv:hep-th/9812073].
}
\refs{\GrossTU , \GrossHU ,\MooreDK , \CordesFC}.
Here we construct a class of non-compact Calabi-Yau threefolds for which the
$Z_{brane}$ is equivalent to a topologically twisted 4 dimensional ${\cal N}=4 $ supersymmetric $U(N)$
Yang-Mills theory which in turn can be mapped to the two dimensional
bosonic Yang-Mills on a two dimensional torus.  We thus
end up identifying the topological string theory encountered
in the large $N$ expansion of Yang-Mills theory with that 
of topological A-model on a local Calabi-Yau threefold.  
This is consistent with the observations of similarities
between large $N$ `t Hooft expansion of 2d Yang-Mills
and topological strings on 3-folds.  For example it was
already shown in
\lref\dij{
R.~Dijkgraaf,
``Chiral deformations of conformal field theories,''
Nucl.\ Phys.\ B {\bf 493}, 588 (1997)
[arXiv:hep-th/9609022].
}
 \dij\ that the large $N$
expansion of 2d Yang-Mills
theory on $T^2$ does satisfy the holomorphic anomaly
equation of 
\lref\bcov{
M.~Bershadsky, S.~Cecotti, H.~Ooguri and C.~Vafa,
``Kodaira-Spencer theory of gravity and exact results for quantum string
amplitudes,''
Commun.\ Math.\ Phys.\  {\bf 165}, 311 (1994)
[arXiv:hep-th/9309140].
} \bcov\ which is a property expected
for topological strings on threefolds.  Moreover we show
using direct computation of the corresponding topological
string that the two sides are equal to all orders in string
perturbation theory.

In fact this example also shows the more general meaning of
\funs :  It is known that for finite $N$ the 2d Yang-Mills
theory does not factorize to a holomorphic square.  This is only
the structure as $N\rightarrow \infty$.  Nevertheless there is a
precise finite answer for finite $N$.  What this means
is that the right hand side of \funs\ is the absolute value
square of the holomorphic function to all orders in string perturbative
expansion (i.e. to all orders in $1/N$) but at finite $N$ this
structure loses its meaning.  In other words there is no sense
to the non-perturbative meaning of $Z_{top}$.  It is only
the ``absolute value squared'' $|Z_{top}|^2$, i.e. a density
operator  which may have a non-perturbative
meaning. In the quantum mechanical correspondence discussed in
\osv\ this would map to the 
statement that the state corresponding to the black
hole is a pure state to all orders in the $1/N$ expansion which becomes mixed
at order $O({\rm exp}(-N))$.  Here $N$ can be viewed as being
proportional to the radius of $AdS_2$.  Moreover we conjecture
that to all orders in the $1/N$ expansion, where $Z_{top}$
and ${\overline Z}_{top}$ are well defined, they can be identified with the two boundaries
of $AdS_2$, thus proposing a {\it single} dual theory for $AdS_2$.

The organization of this paper is as follows:  In section 2
we review the results for the partition function
for $U(N)$ and $SU(N)$ Yang-Mills on $T^2$.  In section
3 we
discuss a particular class of local threefolds (involving
the total space of a direct sum of a line bundle and its inverse on $T^2$)
and show, using the ideas of topological vertex, that the perturbative
amplitudes of topological strings matches the large $N$ expansion
for amplitudes
of the $U(N)$ bosonic Yang-Mills on $T^2$.  In section 4 we use
the ideas in \osv\ to relate the topological string theory
amplitudes to the dual brane theory, which turns out to be
four dimensional ${\cal N}=4$ topologically twisted Yang-Mills theory on a line bundle
over $T^2$.  In section 5 we show that this topologically
twisted theory reduces to a two dimensional topologically twisted
theory on $T^2$ which in turn is equivalent to 2d bosonic Yang-Mills
on $T^2$, thus completing the circle of ideas. In section 6 we
discuss some implications for the black hole physics.

\newsec{Yang-Mills on the two dimensional torus}

Two dimensional Yang-Mills theory was solved in
\lref\RusakovRS{
B.~E.~Rusakov,
``Loop Averages And Partition Functions In U(N) Gauge Theory On Two-Dimensional
Manifolds,''
Mod.\ Phys.\ Lett.\ A {\bf 5}, 693 (1990).
}
\lref\FineZZ{
D.~S.~Fine,
``Quantum Yang-Mills On The Two-Sphere,''
Commun.\ Math.\ Phys.\  {\bf 134}, 273 (1990).
}
\lref\BlauMP{
M.~Blau and G.~Thompson,
``Quantum Yang-Mills theory on arbitrary surfaces,''
Int.\ J.\ Mod.\ Phys.\ A {\bf 7}, 3781 (1992).
}
\lref\MigdalZG{
A.~A.~Migdal,
``Recursion Equations In Gauge Field Theories,''
Sov.\ Phys.\ JETP {\bf 42}, 413 (1975)
[Zh.\ Eksp.\ Teor.\ Fiz.\  {\bf 69}, 810 (1975)].
}
\lref\WittenWE{
E.~Witten,
``On Quantum Gauge Theories In Two-Dimensions,''
Commun.\ Math.\ Phys.\  {\bf 141}, 153 (1991).
}
\refs{\MigdalZG , \RusakovRS ,\FineZZ , \WittenWE ,\BlauMP}.  Moreover this theory was
studied at large $N$ beginning with
\lref\RuddTA{
R.~E.~Rudd,
``The String partition function for QCD on the torus,''
arXiv:hep-th/9407176.
}
\lref\MinahanNP{
J.~A.~Minahan and A.~P.~Polychronakos,
``Equivalence of two-dimensional QCD and the C = 1 matrix model,''
Phys.\ Lett.\ B {\bf 312}, 155 (1993)
[arXiv:hep-th/9303153].
}
\lref\DouglasWY{
M.~R.~Douglas,
``Conformal field theory techniques in large N Yang-Mills theory,''
arXiv:hep-th/9311130.
}
\lref\WittenXU{
E.~Witten,
``Two-dimensional gauge theories revisited,''
J.\ Geom.\ Phys.\  {\bf 9}, 303 (1992)
[arXiv:hep-th/9204083].
}
\GrossTU\ from a number of viewpoints
\refs{ \GrossHU , \MooreDK ,\CordesFC ,\MinahanNP ,\DouglasWY , \RuddTA}.  Here we will only need the results
for the partition function of $U(N)$ and $SU(N)$ Yang-Mills
theory on $T^2$.  Consider the $2d$ Yang-Mills action for group
$G$:
$$S={1\over 2 g_{YM}^2}\int_{T^2} [Tr F^2 +\theta Tr F] $$
where the latter term is non-zero only for $U(N)$.
For simplicity of expression we often take the area of $T^2$ to
be $1$.  The area dependence can be restored by dimensional
analysis ($g_{YM}^2\rightarrow g_{YM}^2 A$).  The partition function
for this theory can be obtained as a sum over representations $R$:
$$\sum_R {\rm exp}(-{1\over 2}g_{YM}^2C_2(R)+i \theta C_1(R))$$
For $SU(N)$ the representations $R$ are given by $2d$ Young
diagrams.  Consider a Young diagram where the $k$-th row
has $n_k$ boxes with $k=1,...,N$, where $n_i\geq n_j$ for $i>j$.
In this case the $C_2(R)$ is given by
$$C_2(R)=\sum_{i=1}^N n_i(n_i+N+1-2i)-{n^2\over N}$$
where 
$$n=\sum_{i=1}^N n_i.$$
  For the $SU(N)$ theory the $C_1(R)=0$.
The representations of the $U(N)$ theory can be obtained
from those of the $SU(N)$ theory by noting the fact that 
\eqn\core{U(N)=SU(N)\times U(1)/Z_N.}
  In other words, by decomposing
the representation in terms of a represention $R$ of $SU(N)$ and a charge $q$
of $U(1)$ we have $q=n+Nr$ reflecting the relation \core .

For the $U(N)$ representation the casimirs are given by
\eqn\rela{C_2(R,q)=C_2(R)+q^2/N}
$$C_1(R,q)=q$$

At large $N$ the $SU(N)$ partition functions splits
up essentially to a product of a holomorphic and an anti-holomorphic function
inolving certain chiral blocks $Z_{\pm}$ of the $U(N)$ theory.  We have
$$Z_+=\sum_{R_+}{\rm exp}(-{1\over 2}g_{YM}^2 C_2(R_+)
+i\theta |R_+|)$$
$$Z_-=Z_+^*$$
where by the sum over $R_+$ we mean summing over all Young diagrams
(i.e. labelling representations of $SU(\infty)$ ) and $|R_+|$ denotes
the total number of boxes.  We define a new casimir $\kappa (R_+)$ by
$$C_2(R_+)=\kappa (R_+) +N |R_+|$$
which is given by
$$\kappa(R_+)=\sum_{i=1}^{\infty} n_i(n_i+1-2i).$$
Note that $\kappa(R_+)$ does not have any explicit
dependence on $N$.  We can write $Z_+$ as
$$Z_+=\sum_{R_+}{\rm exp}(-{1\over 2}g_{YM}^2 \kappa(R_+)-t|R_+|)$$
where 
$$t={1\over 2} Ng_{YM}^2-i\theta.$$
In terms of these chiral blocks the $SU(N)$ partition function
at large $N$ takes the almost factorized form
$$Z_{SU(N)}={\rm exp}[{g_{YM}^2\over 2N}(\partial_t -{\overline
\partial_t})^2] Z_+(t)Z_-({\overline t})$$
where we set $t={\overline t}$ after taking the derivative.  This arises
by splitting the 2d  Young diagrams as excitations with columns with
few boxes (giving $R_+$) combined with columns of order of $N$ boxes
(whose complement gives $R_-$).

It is also useful for us to expand the $U(N)$ partition function
at large $N$.  In this case we have an extra sum over the $U(1)$
charges. Using the relation \rela\ we can relate this to the $Z_{SU(N)}$
as follows:  Let $n_{\pm}$ denotes $|R_{\pm}|$.  Then the total number
of boxes of the $SU(N)$ representation is $Nl_-+(n_+-n_-)$ where $l_-$ is the number of boxes
of the first row of $R_-$.  Since the $U(1)$ charge $q$ is equal
to the number of boxes of the Young diagram mod $N$ we have
\eqn\wq{q=Nr+N l_-+(n_+-n_-)=Nl +(n_+-n_-)}
where we have defined a new variable $l=r+l_-$.  Here
$l$ runs over arbitrary positive and negative integers.  We thus have to add
to partition function of $SU(N)$ a multiplicative term
$${\rm exp}(-g_{YM}^2 q^2/2N)={\rm exp}[{-g_{YM}^2(Nl+n_+-n_-)^2\over
2N}]$$
Note that this multiplicative term cancels the $g_{YM}^2(n_+-n_-)^2/2N$
term which mixes the two chrial block and instead gives
$${\rm exp}(-g_{YM}^2[l(n_+-n_-)]-{1\over 2}Nl^2 g_{YM}^2)=
{\rm exp}(g_{YM}^2l(\partial_t-{\overline \partial_t})-
{1\over 2}Nl^2 g_{YM}^2)$$
This acting on the two blocks just shifts the arguments together
with the extra prefactor
$$ {\rm exp} [{-1\over 2}(t+{\overline t})l^2]\
Z_+(t+lg_{YM}^2)Z_-({\overline t}-lg_{YM}^2)$$

We also have to add the $i\theta q$ term to this.  In order to do
this note that if we do not set $t={\overline t}$ the above
expression would have given $i\theta (n_+-n_-)$.  This is almost $q$
given by \wq .  To correct it we need to also add an extra factor
$${\rm exp}[-N(t-{\overline t})l/2)]={\rm exp}[
{-1\over g_{YM}^2}(t+{\overline t})(t-{\overline t})l/2]=
{\rm exp}[{-1\over 2 g_{YM}^2}(t^2-{\overline t}^2)l]$$

Note that the prefactors can be neatly combined as follows:
Consider a modified 
$$Z_+(t)\rightarrow Z_+(t) \cdot {\rm exp}[{-t^3\over 3!g_{YM}^4}]$$
It is also natural to further multiply $Z_+(t)\rightarrow
Z_+(t) \cdot {\rm exp}(t/24)$ (this will give the usual prefactor
for the one loop term for the Dedekin eta function).
Similarly it is natural to add a `casimir' term to the $U(N)$
Yang-Mills:
$$Z_{U(N)}\rightarrow {\rm exp}[-{t^3+{\overline t}^3\over 3!g_{YM}^4}
+{t+{\overline t}\over 24}] Z_{U(N)}$$
This amounts to adding a term $N(N^2-1)/12$ term to the $C_2(R)$
(when $\theta =0$)
which is natural in the fermionic formulation of 2d Yang-Mills
\refs{\MinahanNP , \DouglasWY}.  In the topological string context the
$N^3$ terms corresponds to the classical amplitude at genus
0 and the $N$ term corresponds to the classical contribution
at genus 1.
Then we have the full $U(N)$ partition function as being
given at large $N$ by
\eqn\etr{ Z_{U(N)}= 
\sum_{l=-\infty}^{+\infty} Z_+(t+lg_{YM}^2)Z_-({\overline t}-lg_{YM}^2)}
It is this factorized form which we will need to make contact
with the predictions of \osv .  Note that non-perturbatively,
i.e. for finite $N$,
the expression \etr\ would not be true.  The full answer
of course exists for finite $N$ but does not take a factorized form
as is suggested in this perturbative version.  Later we will
interpret the sum over $l$ as a sum over RR-fluxes on $T^2$ 
in the type IIA
setup.

\newsec{Local Elliptic Threefold and $U(N)$ Yang-Mills on $T^2$}

Topological A-model amplitudes on local toric threefolds can be effectively
computed using topological vertex 
\lref\akmv{
M.~Aganagic, A.~Klemm, M.~Marino and C.~Vafa,
``The topological vertex,''
arXiv:hep-th/0305132.
}
\akmv.
This involves using the toric graph of the threefold as a Feynman
diagram with cubic vertices, with edges labeled by arbitrary
$2d$ Young-diagrams.  Each edge has a `Schwinger time' $t_i$ 
(which corresponds to K\"ahler classes of the 3-fold) and a 
propagator factor
given by $e^{-t_i|R|}$ where $|R|$ is the number of boxes
of the Young diagram $R$.  In addition each oriented edge can be 
identified
with a primitive 1-cycle $v$ of a two torus.  
With a given choice of 
basis this can be represented by two relatively prime integers
$(p,q)=v$.  In addition each edge comes with a choice of a `framing' 
$f$ which is another choice of a 1-cycle of $T^2$ with the property
that $(f,v)$ forms a good basis for the integral 1-cycles of $T^2$.
This means $f\cap v=1$.  Note that $f$ is unique up to an
integer shift $m$ times $v$
$$f\rightarrow f+mv$$

For any three edges incoming to a vertex we have
$$\sum_{i=1}^3v_i=\sum_{i=1}^3 (p_i,q_i)=(0,0)$$
The interactions are cubic and given
by a vertex $C_{R_1R_2R_3}$.   More precisely
the vertex does depend on the choice of the three framings $f_i$.  
Changing the framing $f_i$ by an integer amount $f_i\rightarrow
f_i+m_iv_i$ changes
$$C_{R_1R_2R_3}\rightarrow q^{\sum_i m_i \kappa (R_i)/2}C_{R_1R_2R_3}$$
where $q=e^{-g_s}$ and $\kappa (R)$ denotes a second casimir of $R$ 
with the unique
property that $\kappa (R^t)=-\kappa(R)$, where $R^t$ denotes the
Young diagram which is the transpose of $R$.

  Fixing a canonical framing and gluing the vertices with compatible
framing leads to arbitrary toric threefolds.  It is possible to choose
different framings on each edge without affecting the geometry as long
as the framing on both vertices are changed in opposite directions.  The
effect of framing would then cancel.  If however one changes the
relative framing between vertices which are being glued together one
changes the intersection numbers of the Calabi-Yau.  This
can be inferred for example from 
\lref\iqbal{A. Iqbal, to appear}
\refs{\akmv, \iqbal}.  Let $C_2$ denote the two cycle
defined by the edge and $C_4$ define a 4-cycle whose projection on toric
base contains the same edge.  Then changing the relative framing of the
edge by $m$ units affects the intersection of $C_2$ and $C_4$ by 
$$\Delta(C_2\cap C_4)=m$$ 
There is a sense of orientation here.  In particular there are
two natural toric 4-cycles containing $C_2$, given by
one of the two faces bordering the edge.  Let us denote
them by $C_4,C_4'$. $C_4$ is determined by the direction
of the framing vector $f$.  For $C_4'$ we have
$$\Delta(C_2\cap C_4')=-m.$$

The elements involved in the definition of the topological string
amplitudes for toric threefold are very similar to that of 2d Yang-Mills\foot{
This has been previously observed by
R. Gopakumar and M. Marino \ref\rgmm{R. Gopakumar and M. Marino
(unpublished), 2003.} in the context of finding a closed string dual to
$2d$ Yang-mills on the sphere.  See also the related work
\ref\deh{S. de Haro and M. Tierz, ``Brownian Motion, Chern-Simons
Theory, and 2d Yang-Mills,'' to appear.}.
}.  Here we wish to make contact with 2d $U(N)$
Yang-Mills on $T^2$, whose chiral partition function is given,
as discussed in the previous section, by 
$$Z^{YM}_+=\sum_R Q^{C_2(R)/2}\ {\rm exp}(i\theta C_1(R))$$ 
where
$Q=exp[-g_{YM}^2A]$ where $A$ is the area of torus and $g_{YM}$ is the
Yang-Mills coupling and $\theta$ is the theta angle in 2d.  For
simplicity of notation we set $A=1$ (by dimensional analysis
it can always be restored).  This is a very similar object to the
propagator of the topological strings, where $C_2(R)$ and ${\kappa (R)}$
differ by terms involving the first casimir $C_1(R)=|R|$.  In fact this
can also be rewritten as discussed in the last section as
$$Z^{YM}_+=\sum_R Q^{\kappa (R)/2} {\rm
exp}(-{1\over 2}Ng_{YM}^2+i\theta)|R|$$ 

Let us define 
$$t_{YM}={1\over
2}Ng_{YM}^2-i \theta$$ 
Then we have 
\eqn\ffor{Z^{YM}_+=\sum_R
Q^{\kappa(R)/2}{\rm exp}(-t_{YM}|R|).} 
This is very similar to the
expressions of topological strings on toric threefolds and we ask
whether or not it corresponds to a particular threefold?  The fact that
there is a trace over all representations suggests that we glue an edge
to itself, giving a `loop' for the toric Feynman diagram. This is only possible if the plane of toric diagram is
periodically identified. It is possible to interpret this
geometrically as was done in \akmv\ and 
\ref\HollowoodCV{
T.~J.~Hollowood, A.~Iqbal and C.~Vafa,
``Matrix models, geometric engineering and elliptic genera,''
arXiv:hep-th/0310272.
}. 
In particular if we just consider gluing an edge to itself we obtain a
torus whose Kahler class is given by the length of the edge $t$.  Note
that this is consistent (up to classical polynomial terms in $t$ that
topological vertex does not compute) with the fact 
that 
$$\sum_R exp(-t
|R|)=1/\prod_{n=1}^{\infty} (1-e^{-nt})$$ 
and that the topolgoical string amplitude on ${\bf C}\times {\bf 
C}\times 
T^2$ is given by the inverse of Dedekind eta function (assuming
that euler character of ${\bf C}^2$ is one) \ref\BershadskyTA{
M.~Bershadsky, S.~Cecotti, H.~Ooguri and C.~Vafa,
``Holomorphic anomalies in topological field theories,''
Nucl.\ Phys.\ B {\bf 405}, 279 (1993)
[arXiv:hep-th/9302103].
}.

To obtain the full structure of the 2d Yang-Mills \ffor\ we also
need to include framing.  In particular before gluing
the vertex back to itself we perform a change of framing by
$m$ units and then glue.  In this way we obtain
$$Z_{top}=\sum_R q^{m\kappa(R)/2}{\rm exp}(-t |R|)$$

We see that $Z_{top}=Z^{YM}_+$ provided that we identify
$$Q=q^m \rightarrow g_{YM}^2=mg_s$$
$$t_{YM}=t \rightarrow {1\over 2}Ng_{YM}^2-i\theta =t\rightarrow
{1\over 2}Nmg_s-i \theta =t$$
Note that the classical triple intersection number for this
geometry would lead to the natural prefactor we introduced
for the $U(N)$ Yang-Mills theory if we assign it to be
$${F_0(t)\over g_s^2}+F_1(t)=-{t^3\over 3! m^2 g_s^2}+{t\over 24}=-{t^3\over 3! g_{YM}^4}+{t\over 24}.$$
Due to the non-compactness of the geometry this is a bit difficult
to apriori justify; nevertheless it is gratifying to see that
the Yang-Mills Casimir prefactors have a natural geometric
interpretation in terms of a choice of triple intersection number
(entering $F_0$) and the second chern class (entering $F_1$).

We next turn to what is the non-compact threefold geometry
defined by this framing operation.

\subsec{The elliptic threefold geometries}
As discussed before the framing operation
affects the interesction numbers.  To begin with we can view
the local geometry before twisting by the framing factor, as a direct sum of two trivial
line bundles on $T^2$:
$${\cal O}(0)\oplus {\cal O}(0)\rightarrow T^2$$
The two faces of the toric digram correspond to the two cycles
$C_4$ and $C_4'$ which in this case are both isomorphic to
$$C_4=C_4'={\cal O}(0)\rightarrow T^2$$
Note that 
$$C_4\cap T^2=0$$

After twisting by the framing the two line bunldes have changed in such a way that their
intersection numbers with $T^2$ change to $m$ and $-m$.  Moreover their line bundles
are canonically inverse of one another (for the threefold to have
trivial canonical line bundle).  We will now show that this implies that
the local threefold is given by 
$${\cal L}^{-m}\oplus {\cal
L}^m\rightarrow T^2$$ 
where ${\cal L}^m$ is a degree $m$ line bundle over $T^2$.  This is unique up to tensoring by a flat bundle.
 ${\cal L}^m$ is characterized by the statement that a holomorphic
section of this bundle has a divisor of degree $m$
on $T^2$ which denotes the zeros of the corresponding holomorphic 
section. 
Moving the $m$ divisors on $T^2$ corresponds to choosing
a different holomorphic section.  ${\cal L}^{-m}$ is the inverse
degree $-m$ bundle.  Each meromorphic section of it will have at least $m$ poles.
  In this context we have
$$C_4={\cal L}^{-m}\rightarrow T^2$$
$$C_4'={\cal L}^m\rightarrow T^2.$$
We will now show that
$$C_4\cap T^2=m.$$
To show this deform $T^2$ to ${T^2}'$ by using a holomorphic
section of ${\cal L}^m$.  As long as we are away from zeros of
this section this does not intersect $C_2$.  The zeros of the 
holomorphic section are precisely where it intersects $C_4$
transversally.  Thus the intersection number is $m$ because there
are $m$ zeros for the holomorphic section.   Note that
this uniquely defines the geometry of the normal bundles.
We have thus shown that the local threefold corresponding
to the case where $Z_{top}=Z_{YM}$ is given by the sum of a degree $m$ 
line
bundle and its inverse on $T^2$.  This local geometry
we shall call $E_m$:
$$E_m={\cal L}^{-m}\oplus {\cal L}^m \rightarrow T^2.$$

\newsec{Black Hole and Elliptic Threefold}

Recently it was suggested in \osv\
that there is a deep connection between 4d BPS black holes and
topological strings, namely the partition function of 4d black
holes is given by the square of the topological
string partition function:
$$Z_{BH}=|Z_{top}|^2$$
where the moduli of the Calabi-Yau is fixed by the attractor
mechanism.  Moreover 
$$Z_{BH}=Z_{brane}$$
is the partition function on the brane defining the black hole, which 
leads to
$$Z_{brane}=|Z_{top}|^2.$$
This is roughly the structure we have here.  Namely if somehow
the brane theory is equivalent to $U(N)$ 2d Yang-Mills, since
$$Z^{YM}\sim Z^{YM}_+Z^{YM}_-=Z_{top}{\overline Z_{top}}$$
we would obtain $Z_{brane}=|Z_{top}|^2$.
This is almost the structure we found for $U(N)$ 2d Yang-Mills
theory at large $N$.  However we found in addition an extra
sum over integer $l$:
$$Z^{YM}=\sum_l Z^{YM}_+(t+lg_{YM}^2)Z^{YM}_-({\overline t}-lg_{YM}^2)$$
This extra sum was related to the $U(1)\subset U(N)$ charge.
Viewed from the topolgoical string side this would lead to the
statement
\eqn\akhar{Z_{brane}=\sum_l Z_{top}(t+mlg_s){\overline Z_{top}}({\overline t}-mlg_s)}
We could have restricted the sum to a fixed $l=0$ and 
interpret it as `freezing' of the $U(1)$ center of mass dynamics
of the black hole.  However, given that the non-perturbative
completion of the $U(N)$ theory requires this sum, it is important
to better understand the physical meaning of this sum
for the black hole partition
function.  Our local geometry $E_m$ is to be viewed as the attractor
fixed point for the black hole geometry. However the full
partition function of the theory should include a sum over all finite
energy configurations, and in this case this includes the sum
over $RR$ flux on the compact cycle $T^2$.  Note that this can
preserve the same amount of supersymmetry as the BPS black hole
and it would be natural to expect that it enters the full
black hole partition function.
It is also natural to consider turning on RR-flux on 2-cycles.
As discussed in 
\lref\adkmv{
M.~Aganagic, R.~Dijkgraaf, A.~Klemm, M.~Marino and C.~Vafa,
``Topological strings and integrable hierarchies,''
arXiv:hep-th/0312085.
} \adkmv\ this has the effect
of shifting
$$Z_{top}\rightarrow Z_{top}(t+lmg_s)$$
where the coefficient $m$ enters here because of the intersection
of $T^2$ with $C_4$ as was explained in \adkmv . The same
reasoning as in \adkmv\ would lead to the statement that for
anti-topological theory ${\overline Z_{top}}\rightarrow 
{\overline Z_{top}}({\overline t}-lmg_s)$.  This means that the
dynamics of the $U(1)$ sector mixes with the RR 2-form flux
(which is natural from the viewpoint of the CS coupling of the
$U(1)$ to the gravitational sector).  We thus end up with a simple 
interpretation of \etr\ as the {\it full} partition function
for the black hole background including a sum over RR fluxes through
$T^2$.  Thus \akhar\ is consistent with the results of \osv\ which focused
on the zero flux sector $l=0$.

\subsec{Setting up the map between 4d and 2d Yang-Mills}

To complete the circle of ideas we need to show that the brane
gauge theory which is the four dimensional
${\cal N}=4$ topologically twisted $U(N)$ Yang-Mills theory 
on $C_4$ is equivalent to bosonic 2d $U(N)$ Yang-Mills theory on $T^2$
with
$$g_{YM}^2=mg_s\qquad \theta_{YM}=\theta .$$

The relation between the charges and the moduli in the type IIA setup is
given by \osv 
$$t^i=2\pi i X^i/X^0$$
$$g_s=4\pi i/X^0$$
where $i=1,...,h^{1,1}$ 
and 
$$X^I=p^I+i{\phi^I\over \pi }$$
where $p^i$ denotes the magnetic charge (given by D4 branes
wrapped over 4-cycles) dual to the electric charges (given by
the D2 branes wrapped on dual 2-cycles) and $\phi^i$ are the chemical
potentials for the electric charge.  Moreover $p^0$ denotes the number
of D6 branes and $\phi^0$ is the chemical potential for the 0-branes.

Even though the main discussion in \osv\ was in the
context of compact Calabi-Yau's, it is natural to expect
that this relation continues to hold (with suitable
boundary conditions) also
for the case of non-compact Calabi-Yau.  This is natural to expect
because the right hand side does make sense for the non-compact
Calabi-Yau's.  In fact this can be viewed as a limit of the
considerations of \osv\ where we take a large volume limit
of a compact Calabi-Yau and focus on a particular sector
of the theory localized near some compact cycles. 

So let us consider our elliptic threefold $E_m$.  There is
only one compact 2-cycle $T^2$ which we identify with an electric
charge.  Here we only have $X^0$ and $X^1$
and
$$t=2\pi i{ X^1\over X^0}$$

  For the mangetic 
charge dual to the electric 2-cycle charge we can consider the four
cycle $C_4$.  However note that the quantum of magnetic
charge is $m$ times bigger on this cycle because $C_4\cap T^2=m$.
Thus if we consider $N$ D4 branes wrapping $C_4$ it will have
magnetic charge $p^1=Nm$.

To make contact with the parameters defined in the previous
section we see that we have to set
$$g_s={4\pi i \over X^0}={4\pi i\over p^0+i{\phi^0 \over \pi}}$$
which for real $g_s$ implies $p^0=0$, i.e. we have no D6 branes.  We thus set
$$g_s={4\pi^2\over \phi^0}\rightarrow \phi^0={4\pi^2\over g_s} $$
where $\phi^0$ is the chemical potential for the 0-brane.  Similarly
we have 
$$X^1={1\over 2\pi i}X^0t={2\over g_s}({1\over 2}Nmg_s+i\theta)=
Nm+{2i\theta \over g_s}$$
which implies 
$$p^1=Nm \qquad \phi^1={2\pi \over g_s}\theta$$
We now ask what is the gauge theory on the brane?  We have no
D6 branes but we do have $Nm$ magnetic branes.  This is reassuring
as it is naturally quantized in multiples of $m$:  We just take
$N$ D4 branes wrapped around $C_4$ and this gives magnetic charge
$Nm$ as discussed before.  This gives a topologically
twisted ${\cal N}=4$, $U(N)$ gauge theory
on $C_4$.  In fact the twisting is exactly
the one studied in 
\ref\vw{
C.~Vafa and E.~Witten,
``A Strong coupling test of S duality,''
Nucl.\ Phys.\ B {\bf 431}, 3 (1994)
[arXiv:hep-th/9408074].
}. 
All we have to do now is to
induce the corresponding chemical potentials for the D0 and D2 branes.
To induce a $D0$ brane charge we have to use an interaction term
${1\over 4\pi^2} \int_{C_4}{1\over 2} {\rm tr} F\wedge F$
which means that we introduce the term 
$$
{\phi^0\over 4\pi^2} \int_{C_4} {1\over 2}{\rm tr} F\wedge F
={{1\over 2g_s}} \int_{C_4}  {\rm tr} F\wedge F $$
To induce the electric charge corresponding to D2 wrapping
$T^2$ we note that the charge is measured by ${1\over 
2\pi}\int_{C_4} F\wedge k$ where $k $ is the unit volume
2-form on $T^2$.  We thus add the additional term to the action
$$\phi^1 \int {F\wedge k \over 2\pi}=
{\theta \over g_s}\int F\wedge k$$
We thus add to the topologically twisted ${\cal N}=4$ Yang-Mills
on $C_4$ the term
\eqn\acti{
{{1\over 2g_s}} \int_{C_4} {\rm tr} F\wedge F 
+{\theta \over g_s}\int F\wedge k}

All we have to do now is to show that the partition function
of this theory is equivalent to that of 2d $U(N)$ Yang-Mills theory
on $T^2$ with $g_{YM}^2=mg_s$ and $\theta_{YM}=\theta$.
We will show in the next section that, modulo
some plausible assumptions, this is indeed the case.

\newsec{${\cal N}=4$ topological Yang-Mills and 2d Yang-Mills}

The ${\cal N}$=4 theory on the $N$ D-branes wrapping $C_4$ is the twisted
theory of the type studied in \vw .  To see this it essentially suffices
to study the spin content of the 3 complex adjoint fields $T,U,V$.  
In the theory studied in \vw , for K\"ahler manifolds, two of them $T,U$ are ordinary scalars
but one of them $V$ is a section of $(2,0)$ line bundle.  As noted in
\ref\bsv{
M.~Bershadsky, V.~Sadov and C.~Vafa,
``D-Branes and Topological Field Theories,''
Nucl.\ Phys.\ B {\bf 463}, 420 (1996)
[arXiv:hep-th/9511222].
}\ to find the twisted theory and in particular the
spin content of the scalars
we have to look at the normal bundle to our brane, i.e. the normal
bundle to $C_4$.  Four of them correspond to moving in spacetime
and these correspond to scalars on $C_4$.  These
we identify with the $T,U$.  The other one $V$ corresponds to moving
$C_4$ inside the Calabi-Yau and thus this corresponds to a section
of $V\in S({\cal L}^{m})$.  But that is precisely the $(2,0)$
bundle on $C_4$.  To see this note that $T dw dz$ would be well defined
if $w$ denotes the coordinate of ${\cal L}^{-m}$ bundle over $T^2$ and
$z$ denotes the coordinates of $T^2$.

To simplify the analysis of this theory we add mass terms
to the adjoint fields as in 
\ref\witdo{
E.~Witten,
``Supersymmetric Yang-Mills theory on a four manifold,''
J.\ Math.\ Phys.\  {\bf 35}, 5101 (1994)
[arXiv:hep-th/9403195].
}\
and \vw .  This corresponds to addition of a superpotential
$$W=mUV+\omega T^2$$
where as explained in \witdo\ $W$ is a section of $(2,0)$ bundle on $C_4$ 
which is satisfied if $m$ is a constant and $\omega$ is a holomorphic
section of the $(2,0)$ line bundle.  
Addition of this term does not affect the
${\cal N}=4$ topological theory.  The simplification for doing this
is that the theory is simpler with ${\cal N}=1$ and in particular
has a mass gap.  The basic strategy of \vw\ was to study
the symmetry structure of this theory.  The simplest
situation would arise if $\omega$ has no zeroes.  In this case
we have a purely ${\cal N}=1$ theory in the IR and this sturcture
and topological invariance was enough to fix all the parameters
(with the help of a few computations for special examples) \vw .
If $\omega$ has zeroes, it will appear at complex codimension one,
i.e. on `cosmic string' loci in the four manifold.   This would then
give additional contributions to the bulk, coming from the cosmic
strings.
 Again
some general facts about what can appear there and the assumption of Montonen-Olive
duality essentially fixed the answer in \vw .

The case at hand has one major difference with
most of the cases considered in \vw :  For us the four manifold $C_4$
is non-compact.
  For such a case no general strategy
was proposed in \vw\ but some special cases related to the
work of Nakajima on instantons on ALE spaces was studied
as a check of Montonen-Olive conjecture.  A crucial role was
played there by the choice of a flat connection at infinity, which
relates to the choice of the holonomy of the $U(N)$ gauge theory
at infinity, i.e. a map $\pi_1(Boundary)\rightarrow U(N)$.

The approach we follow here is to add the mass term, as was
done in \vw\ for the compact cases, but also use the symmetries
of the non-compact theory to constrain our answer. 
For us ${\omega}$ is a section of ${\cal L}^m$.  This means
that each holomorphic section will have $m$ zeroes
localized on $m$ cosmic strings which span $(z_i,w)$  where $z_i$ denotes
the zero of the section $\omega$ as a function of the point $z$ on $T^2$,
and $w$ denotes the coordinate of the holomorphic line 
bundle ${\cal L}^{-m}$ over each zero $z_i$. First let
us ignore these zeroes--we will return to them after
we have discussed the bulk physics.

The fact that the space is non-compact means that we cannot sum
over all the allowed vacua as in the compact case of \vw\ but
rather we have to select which vacuum we are in.  We take this to be
the vacuum corresponding to preserving the $U(N)$ symmetry and
which classically corresponds to setting $T=U=V=0$.

Since our space is non-compact we have to deal with boundary
conditions.  In particular for each point $z$ on the torus
we have a complex plane $w$ representing the fiber of ${\cal L}^{-m}$.
For each such plane we have to specify the boundary condition.
Since the boundary of the complex plane is a circle, all we have
to do is to specify the holonomy $U$ at a circle at infinity. Let
for each $z$,
$\Phi(z)$ denote the generator of this holonomy, i.e., $U=e^\Phi$.
In other words in a suitable gauge
\eqn\wh{\Phi(z) =\oint_{S^1_{z, |w|=\infty}} A.}
What can be said about the field configurations as we come inwards on each fiber?
We note that $C_4$ enjoys a $U(1)$ symmetry corresponding
to phase rotations of the line bundle ${\cal L}^{-m}$:
$$(z,w)\rightarrow (z, e^{i\theta} w)$$
We will make the assumption that the path integral
for the topologically twisted Yang-Mills can be localized to
$U(1)$ invariant configurations. To argue why this can be assumed
note that the K\"ahler metric on each fiber of $C_4$ can be viewed as
$d|w|^2\wedge d\theta$ which is a half-line times a circle.  That we
can restrict the configurations of the supersymmetric theory to be
independent of $\theta$ is plausible as is
familiar from Witten's index $Tr(-1)^F {\rm exp}(-\beta H)$: The
index is independent of $\beta$ and considering $\beta \rightarrow 0$
gives time independent field configurations.  The main difference in our
case is that our space is a half-line and so the configurations can
become somewhat singular near $|w|=0$.  We will assume that the nature
of the singularity is mild and compatible with the $\theta$ invariance
for $|w|>0$.  In particular being $\theta$ independent means
that for each fiber (in a suitable gauge)
$$\int_{fiber} F_{w\overline w}(z,w)dwd{\overline w}=\Phi(z)$$
Note that this circle invariance has effectively reduced
our ${\cal N}=1$ gauge theory on $C_4$ to a topologically
twisted $(2,2)$ supersymmetric
gauge theory on $T^2$.  Under this reduction the observables in
\acti\ map to
$$ \int_{T^2} [{1\over g_s}{\rm Tr} F\Phi +{\theta\over g_s}
{\rm Tr} \Phi]$$

So far we have ignored the $m$ zeroes of $\omega$. At these
points $T$ becomes massless and so integrating $T$ out for the effective
$(2,2)$ theory on $T^2$ should introduce some topologically
invariant point-like
observables on the effective 2d theory at these points.  The 
topologically invariant point-like observables at a point
$z$ are given by
$Tr \Phi^r(z)$ for some positive integer $r$.  We will now
argue that only the $r=2$ term is generated and with strength $1/2g_s$.
In particular we will now show that the term
\eqn\exco{\sum_{i=1}^m{1\over 2g_s}{\rm Tr} \Phi^2(z_i)}
is generated.  We will argue this by consistency as follows:  
When we desribe the field $\Phi$ as the holonomy about the circle
at the infinity of the ${\cal L}^{-m}$ fiber, this assumes a trivialization
of this circle bundle.  Let $1/\vartheta(z,\tau)$ denote a hololomorphic section
of the ${\cal L}^{-m}$ bundle where $\vartheta(z,\tau)$ is a suitable
theta function.  This section will have $m$ poles at $z_i$,
the zeroes of $\vartheta$.  We can use
the argument of $1/\vartheta$ to define a trivialization of the
circle bundle on $T^2-\{z_i\}$.  This gives a global definition
of holonomy $\Phi$ for $T^2-\{ z_i\}$.  We then need to glue
this to the circle bundles over each small neighborhood $U_i$ of
$z_i$.
  However in this gluing we will need to do a `surgery' because
of the poles of $1/\vartheta$:
The circle on $T^2$ going around each $z_i$ is identified with the
circle at infinity over $z_i$.  Let $S^1_i$ denote an inifinitesimal
circle on $T^2$ going around $z_i$,  and let $S^1_{\infty i}$
denote the circle over $z_i$ from the line bundle.  Then we have
to identify these two classes:
$$[S^1_i]=[S^1_{\infty i}].$$
If we consider the holonomy around this circle we learn that
$$\int_{neighborhood\ z_i} F(z)=\int_{S^1_i} A=\int_{S^1_{\infty i}}A=\Phi (z_i)$$
This is exactly consistent with the insertion of the term $
{1\over g_s}{\rm Tr}\Phi^2(z_i)$, because from the term ${1\over g_s} {
\rm Tr}\Phi dA$ we see that $A$ and $\Phi$ are conjugate variables so if
we consider
$${1\over 2g_s}\langle ...\oint_{z_i} A \quad {\rm Tr}\Phi^2(z_i) ...\rangle = \langle ...
\Phi(z_i) ...\rangle$$
reproducing the above geometric fact.   Summing up all these terms
for each point we obtain the term given in \exco .

Since the point observables are independent of which point
they are inserted we can put them at any point and integrate over them with the unit area.
We have thus ended up with the action
$$ \int_{T^2} [{1\over g_s}{\rm Tr} F\Phi +{\theta\over g_s} 
{\rm Tr} \Phi +{m\over 2g_s}{\rm Tr \Phi^2}]$$
As argued in \WittenXU ,
this topologically twisted theory 
is equivalent to the bosonic 2d Yang-Mills theory
upon integrating out $\Phi $ and the fermions.  In particular we get
$$-\int_{T^2} {1\over mg_s} [{1\over 2} {\rm Tr} {F^2}+\theta {\rm Tr}F]$$
Leading to the identification
$$g_{YM}^2=mg_s\qquad \theta_{YM}=\theta.$$
This concludes what we wished to show.

\lref\BalasubramanianKQ{
V.~Balasubramanian, A.~Naqvi and J.~Simon,
``A multi-boundary AdS orbifold and DLCQ holography: A universal holographic
description of extremal black hole horizons,''
arXiv:hep-th/0311237.
}

\newsec{Implications/speculations for the black hole physics}

It was conjectured in \osv\ that a relation of the form
$Z_{BH}=|Z_{top}|^2$ exists for all BPS black
holes in 4 dimensions. 
Note that our geometry is given by $AdS_2 \times X$
where $X=S^2\times CY$.  What is the meaning of this squared structure
$|...|^2$?
In this context it is natural to recall a related puzzle \refs{ \StromingerYG ,
\MaldacenaUZ , \BalasubramanianKQ }:
Unlike higher dimensional $AdS_p$ geometries, $AdS_2$ has {\it two}
boundaries.  This has raised the question of whether
we should have one or two $CFT_1$ duals.  In the present case,
as noted in \osv\ the $AdS_2$ geometry should be holographic
dual to the $D4$ brane worldvolume theory, which as we have
noted reduces, for the ground state sector, to the dynamics
of 2d Yang-Mills.  The geometry of $AdS_2$ is reminiscent of
the structure of 2d $U(N)$ Yang-Mills
theory which at $N>>1$ splits to two chiral sectors which
don't talk with one another.  It is natural to identify
one chiral sector with one boundary of $AdS_2$ and the conjugate
sector with the opposite boundary.  
At infinite $N$ the two boundaries don't talk with one another
just as the two chiral sectors of the $U(N)$ Yang-Mills don't couple.
In terms of our discussion we conjecture
that classically (as $N\rightarrow \infty$)
 we identify $Z_{top}$  with the wave function
at the left boundary of $AdS_2$
and ${\overline Z_{top}}$ with the wave function at the
 right boundary. There is
some evidence that this is the correct picture
\ref\dov{R. Dijkgraaf, R. Gopakumar,
H. Ooguri and C. Vafa, work in progress.}, where one identifies
the dynamics near the two fermi surfaces of the fermionic
formulation of 2d Yang-Mills  \refs{\MinahanNP ,\DouglasWY}\ with the two boundaries of $AdS_2$.

However what is the significance
of the lack of the splitting\foot{For a different string theory
application of lack
of this factorization see \ref\MatsuoNN{
T.~Matsuo and S.~Matsuura,
``String theoretical interpretation for finite N Yang-Mills theory in
two-dimensions,''
arXiv:hep-th/0404204.
}.}
of the two chiral sectors at finite $N$ as we have discovered here?
It was noted that $Z_{top}$
should be viewed as a state in a Hilbert space 
\ref\WittenED{
E.~Witten,
``Quantum background independence in string theory,''
arXiv:hep-th/9306122.
}
which would be natural by associating each to one of the
two boundaries of $AdS_2$.  
To gain a better understanding of the mixing let us rewrite the result for
$Z_{BH}$ as
$$Z_{BH}=\langle Z_{top}|Z_{top}\rangle ={\rm Tr} |Z_{top}\rangle
\langle Z_{top} |={\rm Tr} \rho$$
where $\rho$ is a density matrix.
In this context the lack of factorization at finite $N$ means
that in this context there is no $|Z_{top}\rangle$ which satisfies
the above equation.  In other words $\rho$ is {\it not} a pure
state at finite $N$.  Associating this to the vacuum
state of the black hole would suggest that for finite
$N$ the black hole vacuum is not a pure state.
These ideas are currently under investigation \dov .

\vskip 1cm

I would like to thank R. Gopakumar for many interesting
discussions and for collaboration at an early stage of this work.
I would also like to thank R. Dijkgraaf, S. Gukov,
J. Maldacena, A. Neitzke, H. Ooguri, A. Strominger and T. Takayanagi
for valuable discussions.  

This research was supported in part by NSF grants PHY-0244821 and DMS-0244464.

\listrefs

\end